\documentclass{llncs}
\pdfoutput=1
\usepackage[OT1]{fontenc}
\usepackage[numbers,sort]{natbib}
\usepackage[cmintegrals]{newtxmath}
\usepackage[dvipsnames]{xcolor}
\usepackage{amsmath}
\usepackage{bm}
\usepackage{hyperref}

\definecolor{mygreen}{RGB}{0,150,0}
\definecolor{myred}{RGB}{200,0,0}

\usepackage{tikz}
\usetikzlibrary{arrows, shapes, positioning, calc, intersections}

\tikzstyle{replica} = [rectangle, draw=black!60, fill=black!10, very thick, minimum size=5mm]
\tikzstyle{timeline} = [thick,->,>=stealth]
\tikzstyle{message} = [very thick, ->, >=stealth, shorten <=-0.2mm]
\tikzstyle{operation} = [rectangle, very thin, draw=black!70, fill=black!70, minimum height=3mm, text width=0mm, inner sep=0mm]

\usepackage{fixme}
\fxsetup{
  status=draft,
  layout={inline, marginclue},
  theme=color
}

\newsavebox{\usericon}
\savebox{\usericon}{
\definecolor{c383838}{RGB}{56,56,56}
\definecolor{cff8080}{RGB}{255,128,128}
\definecolor{cffffff}{RGB}{255,255,255}
\definecolor{cffccaa}{RGB}{255,204,170}
\definecolor{cd38d5f}{RGB}{211,141,95}

\begin{tikzpicture}[y=0.80pt, x=0.80pt, yscale=-0.300000, xscale=0.300000, inner sep=0pt, outer sep=0pt]
\begin{scope}[shift={(-459.8,-348.74)}]
  \path[draw=c383838,fill=cff8080,even odd rule,line width=0.773pt]
    (509.0300,412.3700) .. controls (481.2000,413.1600) and (459.3600,458.7500) ..
    (469.6600,474.2000) .. controls (479.9600,489.6600) and (540.3800,489.6600) ..
    (550.2200,474.2000) .. controls (560.0700,458.7500) and (537.7800,411.5500) ..
    (509.0300,412.3700) -- cycle;
  \path[draw=c383838,fill=cffffff,even odd rule,line width=0.835pt]
    (488.3100,421.6800) -- (534.7100,421.6800) .. controls (534.7100,421.6800) and
    (535.4500,467.7800) .. (518.9200,467.7800) .. controls (503.3700,467.7800) and
    (488.3100,421.6800) .. (488.3100,421.6800) -- cycle;
  \path[cm={{0.74318,0.0,0.0,0.82978,(176.87,136.32)}},draw=c383838,fill=cffccaa,even
    odd rule,line width=1.200pt] (496.9600,320.7300)arc(0.234:359.766:46.590);
  \path[draw=c383838,fill=cd38d5f,even odd rule,line width=0.979pt]
    (498.6100,391.4500) .. controls (498.6100,391.4500) and (514.7100,408.4700) ..
    (531.2400,407.1500) .. controls (547.8000,405.8200) and (558.9500,398.1000) ..
    (558.9500,398.1000) .. controls (558.9500,398.1000) and (550.4100,351.5600) ..
    (512.2700,350.7900) .. controls (482.4100,350.1700) and (467.7100,378.8300) ..
    (463.3500,398.1000) .. controls (458.9300,417.6200) and (464.9900,417.5500) ..
    (471.4400,415.8800) .. controls (477.8900,414.2200) and (498.6100,391.4500) ..
    (498.6100,391.4500) -- cycle;
  \path[draw=c383838,fill=cffccaa,even odd rule,line width=0.720pt]
    (565.0600,449.3900) .. controls (569.2800,455.0600) and (567.2700,463.7200) ..
    (560.5700,468.7000) .. controls (553.8700,473.6800) and (545.0000,473.1200) ..
    (540.7800,467.4500) -- (540.7200,467.3600) .. controls (538.2600,463.9800) and
    (537.8100,449.4800) .. (542.2000,446.4500) .. controls (547.0300,443.1200) and
    (560.8400,443.7100) .. (565.0600,449.3900) -- cycle;
\end{scope}

\end{tikzpicture}
}

\newcommand{\vis}{\ensuremath{\mathsf{vis}}}

\newcommand{\set}{\mathtt{put}}
\newcommand{\get}{\mathtt{get}}

\newcommand{\grant}{\ensuremath{\mathtt{grant}}}
\newcommand{\revoke}{\ensuremath{\mathtt{revoke}}}
\newcommand{\exec}{\ensuremath{\mathtt{exec}}}

\newcommand{\lastevent}{\ensuremath{\mathit{last\_event}}}

\begin{document}

\title{EPTL - A temporal logic for weakly consistent systems}

\author{Mathias Weber
 \and Annette Bieniusa
 \and Arnd Poetzsch-Heffter}
\institute{University of Kaiserslautern, Kaiserslautern, Germany\\
 \email{\{m\_weber,bieniusa,poetzsch\}@cs.uni-kl.de}}

\maketitle              % typeset the title of the contribution

\begin{abstract}
 The high availability and scalability of weakly-consistent systems attracts system designers.
 Yet, writing correct application code for this type of systems is difficult;
 even how to specify the intended behavior of such systems is still an open question.
 There has not been established any standard method to specify the intended dynamic behavior of a weakly consistent system.
  There exist specifications of various consistency models for distributed and concurrent systems \citep{crdt,shapiro_consistency_2016};
 and the semantics of replicated datatypes like CRDTs\citep{crdt} have been specified in axiomatic and operational models based on visibility relations.

 In this paper, we present a temporal logic, EPTL, that is tailored to specify properties of weakly consistent systems.
 In contrast to LTL and CTL, EPTL takes into account that operations of weakly consistent systems are in many cases not serializable and have to be treated respectively to capture the behavior.
 We embed our temporal logic in Isabelle/HOL and can thereby leverage strong semi-automatic proving capabilities.
\end{abstract}

\section{Introduction}

To improve availability and fault tolerance, information systems are often replicated to several nodes and globally distributed.
In such system scenarios, designers face a trade-off between availability, fault tolerance, and consistency.
To achieve high availability, designers might weaken the consistency constraints between the nodes.
For example, the replicated state might consist of several objects and communication is done by asynchronous message passing for communication.
In weakly consistent systems, we might refrain from making the objects consistent after each operation.
Operations are first applied to the local objects and then asynchronously sent to the other nodes.

In such systems with weak consistency semantics, concurrent modifications of a replicated object can lead to a divergent system state as the order in which updates are applied can differ among the nodes.
To avoid the divergence of the system state, these update conflicts need to be resolved.
One way to solve conflicts is to use CRDTs \citep{crdt}.
The main idea of CRDTs is to leveraging mathematical properties of the data structure and its operations to automatically solve conflicts due to concurrent modifications of the state of a replicated object.

An easy example of a CRDT is a counter with the operations to get the current value of the counter and an increment operation to increment the counter by one.
Instead of reading the value, incrementing it by one and writing the new value back, an increment operation itself is registered by the counter object.
The current value of the counter object on some node $N$ can be computed by adding up all increment operations known to the node.
The origin of the operations is not important for the computation of the value so concurrent increment operations do not conflict with each other.
Counter increments are commutative which means that applying them to the local state is independent of the order in which the operations are received.

We can see each execution of an operation on a node as an event on the particular node.
Each replica sees a different sequence of events, some of which are synchronized with other nodes thereby becoming part of the event trace of both nodes.
The standard notion of time as being linear is known to not work well in weakly consistent systems as described by \citet{lamport1978}.
Instead of assuming linear time, we consider time as a partial order on the events in the system as already proposed by \citet{burckhardt2014}.

Our goal is to have a specification language for properties of weakly consistent systems.
The specification should be independent of the conflict resolution strategy used in the concrete implementation because this strategy partially depends on the required properties.
The topic of specifying weakly consistent systems is an open research question.
LTL\citep{pnueli_temporal_1977} is a classical specification language for dynamic properties of systems.
It is widely used to specify properties of reactive systems.
LTL is known for formulas which are easy to understand as well as its formal foundation.
This specification language is only recently being used to specify weak memory consistency \citep{senftleben16}.
As we will show in Section \ref{sec:LTLnotsuitable}, it can be difficult to capture the concurrent nature and asynchronous communication typical for weakly consistent systems in LTL (and CTL).

Current approaches tend to base the behavior of the weakly consistent system on the specification of the conflict resolution mechanism used \citep{zeller_towards_2016,gotsman_composite_2015,burckhardt2014}.
For many replicated data types like sets and maps, there are multiple possible implementations each with different semantics for concurrent modifications.
We want to decouple the specification of the behavior of the system from the behavior of the data types and want to enable to choose the right implementation based on the required properties described in the system specification.
Our focus is on the understandability of the specification as well as a solid formal foundation.

The paper makes the following contributions:
\begin{itemize}
  \item We show why current temporal logics are not suitable to specify the intended behavior of weakly consistent systems (Section \ref{sec:LTLnotsuitable}).
  \item We present our event-based parallel temporal logic (EPTL) which is based on an abstract execution of the system and allows to express properties on the global partial order of the events of the system taking into account the non-serializability of operations (Section \ref{sec:eptl}).
  \item We present laws that allow to rewrite EPTL formulas while retaining the semantics (Section \ref{sec:laws}).
  \item EPTL is modeled in Isabelle/HOL and all laws are formally verified.
\end{itemize}

\section{Abstract executions}

When specifying the behavior of a weakly consistent system, we need to formalize the behavior of such a system.
To motivate our approach, let us start with an example.

Weakly consistent systems are composed of multiple processes.
Instead of sharing the state directly and protecting concurrent accesses using locks, each process obtains a replica of the shared object and solely interacts with this object.
The values of the replicas are synchronized by asynchronously distributing the operations to all replicas.
A typical data structure used in such systems is a \emph{multi-value register (MVR)}.
This datatype ensures that all written values of concurrent write operations are visible to subsequent read operations.
The $\set$ operations allows to assign a new value to the register, the $\get$ operation allows to access the current state.
Since the result of the $\get$ operation can consist of multiple concurrently written values, the result is a set of values.
This means that if concurrently we have an operation writing the value 1 and one operation writing the value 2, the value of the register after synchronization of the operations is the set $\{1, 2\}$.
Note that this property of multi-value registers usually leads to non-serializable system traces.

When formally specifying the semantics of the multi-value register, we want to abstract away from details concerning communication and process structure.
Following \citet{burckhardt2014}, we model the execution of a weakly consistent system
as an abstract execution.
An abstract execution $A$ consists of a set of events $E$ and a visibility relation $vis \subseteq E \times E$.
The set $E$ denotes the events representing the execution of operations on different nodes of the distributed weakly consistent system.
The events have a unique identity and carry the metadata about the object and operation executed on it as well as information relevant for the specific use case like which subject executed the operation.
The $vis$ relation models the dependency between events.
For two events $e_1$ and $e_2$ if $(e_1, e_2) \in vis$ than $e_1$ can influence the effect of $e_2$.
The local order of events for each process is usually included in the visibility relation.
We require that the visibility relation is irreflexive, transitive and antisymmetric.
This corresponds with cross-object causal consistency as presented in \citep{burckhardt2014}.
In addition, the visibility relation needs to be well-founded so we can talk about the next events in the execution.
The relation can also be depicted in an \emph{event graph} where the nodes of the graph are the events and the edges represent the visibility relation.
Transitive edges are left out for readability.

We annotate the nodes of event graphs with \emph{operation expressions} as follows:

$$
\begin{array}{l}
  op(p_1, \dots, p_n)\\
  op(p_1, \dots, p_n) \Rightarrow \mathit{retval}
\end{array}
$$

The first form describes that an event $e$ represents an execution of operation $op$ with parameters $p_1$ to $p_n$.
If the returned value is important, we denote it as the second form where $op(p_1, \dots, p_n)$ is defined as above and $\mathit{retval}$ represents the returned value.

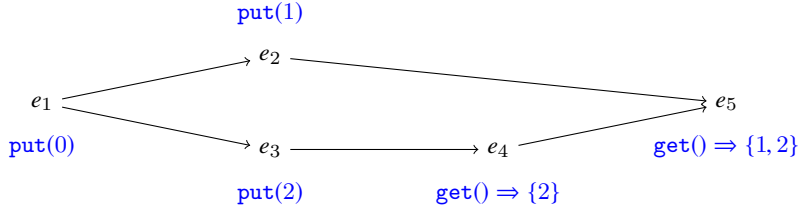
\begin{figure}
  \begin{center}
    \begin{tikzpicture}
     \node (e1) {$e_1$};
     \node[above right = 0.2cm and 2.5cm of e1] (e2) {$e_2$};
     \node[below right = 0.2cm and 2.5cm of e1] (e3) {$e_3$};
     \node[right = 2.5cm of e3] (e4) {$e_4$};
     \node[above right = 0.2cm and 2.5cm of e4] (e5) {$e_5$};
     \draw[->] (e1) -- (e2);
     \draw[->] (e1) -- (e3);
     \draw[->] (e2) -- (e5);
     \draw[->] (e3) -- (e4);
     \draw[->] (e4) -- (e5);
     \node[below = 0.1cm of e1, blue] {$\set(0)$};
     \node[above = 0.1cm of e2, blue] {$\set(1)$};
     \node[below = 0.1cm of e3, blue] {$\set(2)$};
     \node[below = 0.1cm of e4, blue] {$\get() \Rightarrow \{2\}$};
     \node[below = 0.1cm of e5, blue] {$\get() \Rightarrow \{1,2\}$};
    \end{tikzpicture}
  \end{center}
  \caption{Event graph of a multi-value register.}
  \label{fig:mvr}
\end{figure}

In form of an event graph, the example for the multi-value register can be depicted as in Figure \ref{fig:mvr}.
Event $e_1$ corresponds to an initial $\set$ operation, which assigns the single value $0$.
The $\set(1)$ operation of $e_2$ happens concurrently with another operation, $\set(2)$ of $e_3$, that also modifies the state of the register.
Both events are visible to event $e_5$ associated to the $\get$ operation which yields the set $\{1,2\}$ as result.
As the example shows, this abstract execution is only concerned with the partial order of events with respect to the visibility relation; the event graph abstracts away from the details of a specific implementation (e.g. which process executes an operation or how operations are distributed to the other process).

\section{Why LTL and CTL are not suitable}
\label{sec:LTLnotsuitable}

To capture the semantics of weakly consistent systems, we examined the existing logics LTL\citep{pnueli_temporal_1977} and CTL\citep{clarke_design_1981}.
As we are going to show, both these logics are a bad fit when it comes to specifying the semantics of data types such as the multi-value register.

We start with a standard definition of LTL as presented by \citet{Lichtenstein1985}:

$$
\begin{array}{ll}
 (\sigma, j)\models Q                          & \text{ iff } Q \in I(s_j)                                                          \\
 (\sigma, j)\models \neg\varphi                & \text{ iff } (\sigma, j) \not\models\varphi                                        \\
 (\sigma, j)\models (\varphi_1 \vee \varphi_2) & \text{ iff } (\sigma, j)\models \varphi_1 \text{ or } (\sigma, j)\models \varphi_2 \\
 (\sigma, j)\models X \varphi                  & \text{ iff } j+1 < |\sigma| \text{ and } (\sigma, j+1)\models \varphi              \\
 (\sigma, j)\models (\varphi U \psi)           & \text{ iff } \exists k. j\leq k < |\sigma| \text{ and } (\sigma, k) \models \psi \text{ and }  \\
                                               & \qquad \forall i. j\leq i < k \text{ then } (\sigma, i)\models \varphi
\end{array}
$$

with the usual operators defined as follows:

$$
\begin{array}{ll}
 (\sigma, j)\models F \varphi        & \text{ iff } (\sigma, j) \models true ~U~ \varphi                  \\
 (\sigma, j)\models G \varphi        & \text{ iff } (\sigma, j) \models \neg F \neg \varphi               \\
 (\sigma, j)\models \varphi ~W~ \psi & \text{ iff } (\sigma, j) \models G \varphi \vee (\varphi ~U~ \psi)
\end{array}
$$

In this model $\sigma$ is a sequence of states $S$ and $I: S \rightarrow 2^\Pi$ is an evaluation such that $I(s) \subseteq \Pi$ is the set of propositions that are true in $s$.
A computation in this model is a possibly infinite sequence of states such that $\sigma = s_0, s_1, ...$.
The length of $\sigma$ is defined to be the number of states in the sequence if $\sigma$ is finite and $\omega$ otherwise (i.e. the cardinality of the natural numbers).
The $X$ operator defines a strong step such that $(\sigma, j) \models X \varphi$ means that $\varphi$ has to hold in the next step $j + 1$.
The $U$ operator stands for the strong until such that $(\sigma, j) \models \varphi ~U~ \psi$ means that $\psi$ has to hold in the future and all states between than and the current state have to satisfy $\varphi$.

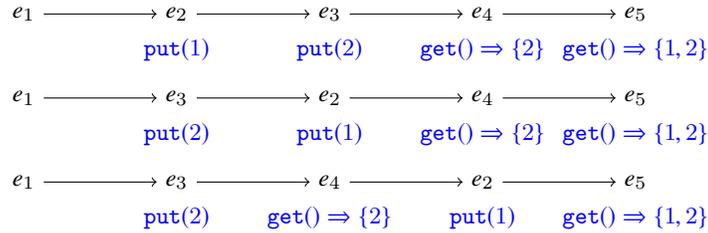
\begin{figure}
  \begin{center}
    \begin{tikzpicture}
     \node                       (e1_1) {$e_1$};
     \node[right = 1.5cm of e1_1] (e2_1) {$e_2$};
     \node[right = 1.5cm of e2_1] (e3_1) {$e_3$};
     \node[right = 1.5cm of e3_1] (e4_1) {$e_4$};
     \node[right = 1.5cm of e4_1] (e5_1) {$e_5$};
     \draw[->] (e1_1) -- (e2_1);
     \draw[->] (e2_1) -- (e3_1);
     \draw[->] (e3_1) -- (e4_1);
     \draw[->] (e4_1) -- (e5_1);
     \node[below = 0cm of e2_1, blue] {$\set(1)$};
     \node[below = 0cm of e3_1, blue] {$\set(2)$};
     \node[below = 0cm of e4_1, blue] {$\get() \Rightarrow \{2\}$};
     \node[below = 0cm of e5_1, blue] {$\get() \Rightarrow \{1,2\}$};

     \node[below=0.7cm of e1_1]   (e1_2) {$e_1$};
     \node[right = 1.5cm of e1_2] (e3_2) {$e_3$};
     \node[right = 1.5cm of e3_2] (e2_2) {$e_2$};
     \node[right = 1.5cm of e2_2] (e4_2) {$e_4$};
     \node[right = 1.5cm of e4_2] (e5_2) {$e_5$};
     \draw[->] (e1_2) -- (e3_2);
     \draw[->] (e3_2) -- (e2_2);
     \draw[->] (e2_2) -- (e4_2);
     \draw[->] (e4_2) -- (e5_2);
     \node[below = 0cm of e2_2, blue] {$\set(1)$};
     \node[below = 0cm of e3_2, blue] {$\set(2)$};
     \node[below = 0cm of e4_2, blue] {$\get() \Rightarrow \{2\}$};
     \node[below = 0cm of e5_2, blue] {$\get() \Rightarrow \{1,2\}$};

     \node[below=0.7cm of e1_2]   (e1_3) {$e_1$};
     \node[right = 1.5cm of e1_3] (e3_3) {$e_3$};
     \node[right = 1.5cm of e3_3] (e4_3) {$e_4$};
     \node[right = 1.5cm of e4_3] (e2_3) {$e_2$};
     \node[right = 1.5cm of e2_3] (e5_3) {$e_5$};
     \draw[->] (e1_3) -- (e3_3);
     \draw[->] (e3_3) -- (e4_3);
     \draw[->] (e4_3) -- (e2_3);
     \draw[->] (e2_3) -- (e5_3);
     \node[below = 0cm of e2_3, blue] {$\set(1)$};
     \node[below = 0cm of e3_3, blue] {$\set(2)$};
     \node[below = 0cm of e4_3, blue] {$\get() \Rightarrow \{2\}$};
     \node[below = 0cm of e5_3, blue] {$\get() \Rightarrow \{1,2\}$};
    \end{tikzpicture}
  \end{center}
  \caption{Possible serializations of the event graph in Figure \ref{fig:mvr}.}
  \label{fig:serializations}
\end{figure}

Let us consider the MVR semantics in LTL.
Because LTL is defined on a sequence of states, formalizations in LTL require to encode the system behavior as some (sequential) state representation.
For the MVR, we first need to compute the sequentializations of the event graph.
Figure \ref{fig:serializations} shows the possible serializations of the event graph depicted in Figure \ref{fig:mvr}.

If we regard the serializations as independent event graphs, none of the executions yields a result for a $\get$ operation that consists of more than one value since none of the $\set$ operations happen concurrently.
To distinguish these serializations from event graphs that represent executions without concurrency, we need to encode the relations between the events in the original event graph into the state we use for LTL.
The approach is similar to the one used by \citet{alur_deciding_2005} in that for each concurrent process we encode a separate state and employ some form of meta-data to capture the visibility relation of the events.
But this does not scale well based on the number of processes in the system.
For typical weakly consistent systems, the number of replicas participating in the system might be in the hundreds, which makes this approach unfeasible.

A second argument against the state encoding is that this encoding usually requires some form of knowledge about the implementation of the data type.
There are multiple CRDTs available that all represent a set.
The implementation only differ in the properties they guarantee.
A specification of the dynamic system properties should not depend on the implementation but instead allow to choose the right implementation to use by showing that a specific implementation of a data type shows the required properties of the system specification.

Another possibility is to encode the partial order of the events into the LTL formulas itself.
We saw that it is not feasible to sequentialize the events thereby getting an exact linear time.
Instead we could use the real-time order of the events (or an approximation thereof) and add additional formulas to capture the visibility of events.
This would allows us to represent the exact dependencies between events without relying on the sequence of states.

The problem with this approach is that the temporal operators do not work on the partial order of events.
Instead they are now based on an approximation of the time which cannot be an exact representation of the actual order.
The formulas yielded by this approach already express the temporal relationships without relying on the temporal operations of LTL and thus make these operators obsolete.
We thereby loose the advantages we wanted to gain by trying to express the required properties using an LTL-like logic.

\paragraph{CTL approach.}
LTL is not right approach for expressing properties of weakly consistent systems mainly because we do not have a linear time.
CTL on the other hand can express properties based on a branching time, which should be a better match for the partial order of events we observe in weakly consistent systems.
With branching time we can also express that multiple different events can be the successor of a single event.
The problem we still face is that in weakly consistent systems, the order of events forms a directed acyclic graph (DAG) instead of a tree.
But CTL is based on a tree-like structure as base for the time in the system.
We need to find a way to transform the DAG of time into a tree of time.
We already know that we cannot sequentialize the events because of problems discussed before.
The only option left is to duplicate events for the transformation.
We look at an example of how this would work:
When transforming the event graph in Figure \ref{fig:mvr} into a tree, the result might look like depicted in Figure \ref{fig:ctl}:

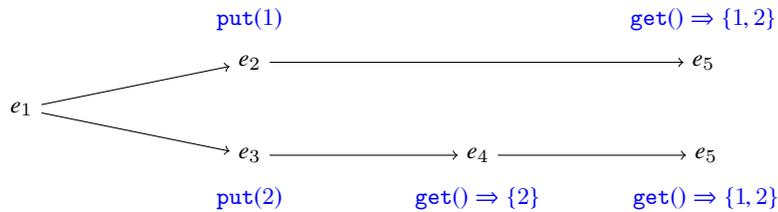
\begin{figure}
  \begin{center}
    \begin{tikzpicture}
     \node (e1) {$e_1$};
     \node[above right = 0.2cm and 2.5cm of e1] (e2) {$e_2$};
     \node[below right = 0.2cm and 2.5cm of e1] (e3) {$e_3$};
     \node[right = 2.5cm of e3] (e4) {$e_4$};
     \node[right = 0.2cm and 5.5cm of e2] (e5_1) {$e_5$};
     \node[right = 0.2cm and 2.5cm of e4] (e5_2) {$e_5$};
     \draw[->] (e1) -- (e2);
     \draw[->] (e1) -- (e3);
     \draw[->] (e2) -- (e5_1);
     \draw[->] (e3) -- (e4);
     \draw[->] (e4) -- (e5_2);
     \node[above = 0.1cm of e2, blue] {$\set(1)$};
     \node[below = 0.1cm of e3, blue] {$\set(2)$};
     \node[below = 0.1cm of e4, blue] {$\get() \Rightarrow \{2\}$};
     \node[above = 0.1cm of e5_1, blue] {$\get() \Rightarrow \{1,2\}$};
     \node[below = 0.1cm of e5_2, blue] {$\get() \Rightarrow \{1,2\}$};
    \end{tikzpicture}
  \end{center}
  \caption{Event tree after transforming the event graph in Figure \ref{fig:mvr}.}
  \label{fig:ctl}
\end{figure}

Since both writes happen concurrently, the resulting value of the $\get$ operation when executing event $e_5$ is $\{1, 2\}$.
Looking at the original DAG, this value can clearly be justified.
But looking at the resulting tree after the transformation, this value at $e_5$ cannot be fully justified.
In the upper branch at $e_2$, only the write of value $1$ is visible, which does not justify the additional value $2$ in the result of $\get$ at $e_5$.
For the lower branch at $e_3$, we can only justify the value $2$, not $1$.
In short, we lost valuable information about the concurrency of events using the DAG-to-tree transformation.

Summarizing the results of the discussion we see that neither LTL nor CTL is suitable to describe dynamic properties of weakly consistent systems.
This leads us to define our own temporal logic based on the ideas of LTL to define dynamic properties of weakly consistent systems directly based on the event graph.

\section{Event-based parallel temporal logic (EPTL)}
\label{sec:eptl}

In this section we present a new variant of temporal logic, namely event-based parallel temporal logic (EPTL).
Instead of being based on possible states of the system, this logic is directly based on events following many previous works \citep{havelund_testing_2001,diekert_pure_2006,thiagarajan_expressively_2002,alur_deciding_2005}.
For an abstract execution $A = (E, \vis)$ we define the partial order $e_1 \leq_A e_2 \equiv e_1 = e_2 \vee (e_1, e_2) \in \vis$.
When $A$ is clear from the context, we simply write $e_1 \leq e_2$.
The satisfaction relation $(A,e) \models \varphi$ is defined recursively over the structure of the formula as follows:

$$
\begin{array}{ll}
 (A, e)\models Q                          & \text{ iff } Q[I](e) \text{ for variable interpretation } I \\
 (A, e)\models \neg\varphi                & \text{ iff } (A, e) \not\models\varphi                                                                                        \\
 (A, e)\models (\varphi_1 \vee \varphi_2) & \text{ iff } (A, e)\models \varphi_1 \text{ or } (A, e)\models \varphi_2                                                      \\
 (A, e)\models EX \varphi                 & \text{ iff } \exists e_1. e < e_1 \text{ and $e_1$ is a minimum wrt $<$ and } (A, e_1)\models \varphi      \\
 (A, e)\models AX \varphi                 & \text{ iff } \forall e_1. e < e_1 \text{ if $e_1$ is a minimum wrt $<$ then } (A, e_1)\models \varphi \\
 (A, e)\models (\varphi ~U~ \psi)         & \text{ iff } \exists e_1. e \leq e_1 \text{ such that } (A, e_1) \models \psi) \text{ and }                                                    \\
                                          & \qquad \forall e_3. e \leq e_3 \text{ such that } (A, e_3)\not\models \varphi \text{ exists $e_2$ such that }\\
                                          & \qquad\qquad e \leq e_2 \text{ and } e_2 \leq e_3 \text{ and } (A, e_2) \models \psi
\end{array}
$$

An interpretation $I$ assigns values to all free variables occurring is an EPTL formula.
$Q[I]$ stands for the proposition $Q$ in which all free variables are replaced by their interpretation according to $I$.
An EPTL formula $\varphi$ is said to be valid if  $(A,e)\models \varphi$ for all interpretations $I$.
An abstract execution $A$ satisfies an EPTL property $\varphi$ written $A \models \varphi$ if all starting events of the abstract execution satisfy $\varphi$.
The starting events of an abstract execution $A$ are all events that are minimal with respect to the partial order $\leq_A$ so which have no predecessor events.

The logical operators $\wedge$ and $\Rightarrow$ can be defined as usual.
The remaining temporal logic operators can be defined as follows:

$$
\begin{array}{ll}
 (A,e)\models F \varphi        & \text{ iff } (A,e) \models true ~U~ \varphi                  \\
 (A,e)\models G \varphi        & \text{ iff } (A,e) \models \neg F \neg \varphi               \\
 (A,e)\models \varphi ~W~ \psi & \text{ iff } (A,e) \models G \varphi \vee (\varphi ~U~ \psi)
\end{array}
$$

The semantics of the $F$ and $G$ operators is as usual:

$$
\begin{array}{ll}
 (A,e)\models F \varphi & \text{ iff } \exists e_1. e \leq e_1 \text{ and } (A,e_1)\models \varphi      \\
 (A,e)\models G \varphi & \text{ iff } \forall e_1. e \leq e_1 \text{ holds that } (A,e_1)\models \varphi \\
\end{array}
$$

The main difference to LTL is that we have two different step operators $EX$ and $AX$ and a different semantics for the until operator $U$ which is tailored to weakly consistent systems.
Because the events in the system are ordered using a partial order, the next step is no longer unambiguous.
Because of branches of concurrent events, a step might address multiple subsequent events.
We want to have the possibility to address either at least one ($EX$) or all ($AX$) events that happen immediately after the current event.
We will use these operators in Section \ref{sec:laws} to define laws that hold for EPTL.
Also the semantics of the until operator $U$ has to be adapted to the partial order.
The semantics is best explained based on the event graph of an abstract execution.

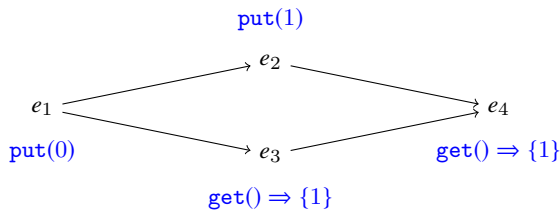
\begin{figure}
  \begin{center}
    \begin{tikzpicture}
     \node (e1) {$e_1$};
     \node[above right = 0.2cm and 2.5cm of e1] (e2) {$e_2$};
     \node[below right = 0.2cm and 2.5cm of e1] (e3) {$e_3$};
     \node[above right = 0.2cm and 2.5cm of e3] (e4) {$e_4$};
     \draw[->] (e1) -- (e2);
     \draw[->] (e1) -- (e3);
     \draw[->] (e2) -- (e4);
     \draw[->] (e3) -- (e4);
     \node[below = 0.1cm of e1, blue] {$\set(0)$};
     \node[above = 0.1cm of e2, blue] {$\set(1)$};
     \node[below = 0.1cm of e3, blue] {$\get() \Rightarrow \{1\}$};
     \node[below = 0.1cm of e4, blue] {$\get() \Rightarrow \{1\}$};
    \end{tikzpicture}
  \end{center}
  \caption{Event graph of an invalid execution for a multi-value register.}
  \label{fig:invalmvr}
\end{figure}

One of the properties of a MVR is that after putting a value into the register the $\get$ operation returns this value until there is a subsequent $\set$ operation.
This property can be expressed in EPTL as the formula
$$G(\set(a) \Rightarrow (a \in \get() ~W~ \set(b)))$$
The proposition $a \in \get()$ is true if the event is an execution of the $\get$ operation and the value $a$ is in the set returned by this operation.
The example execution in Figure \ref{fig:mvr} satisfies this property.
On the other hand, Figure \ref{fig:invalmvr} shows an execution that is not valid since it does not satisfy the property of a MVR.
Event $e_1$ represents an execution of a $\set$ operation of value $0$, which means that future $\get$ operations should return this value until a $\set$ operation of a different value is executed.
Event $e_2$ is such an execution setting value $1$, which justifies the result $\{1\}$ of the $\get$ operation execution represented by $e_4$.
On the other hand, the result of the execution of the $\get$ operation $e_3$ does not satisfy the presented EPTL formula.

Why is this sensible?
The synchronization between the concurrent processes is given by the joins in the event graph.
The events $e_2$ and $e_3$ happen concurrently without information exchange so we cannot assume event $e_2$ to justify that the execution of $\get$ in $e_3$ returns $\{1\}$.
The definition of the until operation is stronger than in previous work \citep{thiagarajan_expressively_2002,alur_deciding_2005,diekert_pure_2006} to be able to express strong properties about weakly consistent systems like the correctness of access control.

\paragraph{Access Control Example}

In this section we want to show an example of properties about weakly consistent applications that can be expressed using EPTL.

One of our starting points was that we wanted to specify exactly what access control in weakly consistent systems means.
Since this is a safety-critical question, we need a specification that is easy to understand and at the same time has a strong semantics on the execution of such a weakly consistent application.
In general, access control is about specifying which operations are permitted to be executed by some subject or user on some object in the system.
In a simple access control system we consider three types of operations:

\begin{itemize}
 \item $\grant(op, s, o)$ gives subject $s$ the right to perform operation $op$ on object $o$
 \item $\revoke(op, s, o)$ takes away the right of subject $s$ to perform operation $op$ on object $o$
 \item $\exec(op, s, o)$ represents the execution of operation $op$ performed by subject $s$ on object $o$
\end{itemize}

Corresponding propositions (e.g. $\grant_P(op, s, o)$) are true for an event $e$ if $e$ represents the execution of the corresponding operation with the given parameters (e.g. $\grant(op, s, o)$).

Based on the given operations, we can define the properties we require from our simple access control system.
We want to start with a default policy that initially no user has the right to execute any operations on the system until an administrative user grants this right to the subject.
To simplify the example, we do not consider the details of rights to perform grant and revoke operations and assume that there is some administrative user in the system that has the right to perform these operations.
The initial policy can be specified in EPTL by the following property:

$$
A \models \neg \exec_P(op, s, o) ~W~ \grant_P(op, s, o)
$$

The dependency between grant and revoke should work like this:
Whenever the right of a subject is revoked, this operation should not be executed until a subsequent grant allows the operation again.
This can be specified in EPTL in the following way:

$$
A \models G(\revoke_P(op, s, o) \Rightarrow AX(\neg\exec_P(op, s, o) ~W~ \grant_P(op, s, o)))
$$

This property both models the semantics of the revoke and grant operations.
A grant operation allows an operation that was previously revoked and a subsequent revoke operation disables the operation for the specified user again.

We see that the specifications are both readable and understandable as well as short.
The strong semantics of the until operator ensures that revoking the right of a user disallows the operation on all future concurrent paths in the event graph.

\section{Laws of EPTL}
\label{sec:laws}

Do the laws that hold for LTL also extend to EPTL?
In the following, we discuss which laws are also applicable for EPTL.
For the rules that hold, we derived proofs in Isabelle/HOL;
for the equalities that do not hold, we explain why they cannot be valid in EPTL due to the partial order imposed on events.

\paragraph{Distributivity.}
We start with the rules of distributivity.
The following rules are proven to be valid rewrites in EPTL:

\begin{eqnarray*}
 EX \varphi \vee EX \psi &\equiv& EX(\varphi \vee \psi)                      \\
 AX \varphi \wedge AX \psi &\equiv& AX (\varphi \wedge \psi)                 \\
 (F \varphi) \vee (F \psi) &\equiv& F(\varphi \vee \psi)                     \\
 (G \varphi) \wedge (G \psi) &\equiv& G(\varphi \wedge \psi)                 \\
 (\varphi U \rho) \wedge (\psi U \rho) &\equiv& (\varphi \wedge \psi) U \rho
\end{eqnarray*}

But there are some laws where only one implication holds.
We will discuss them in detail here.

$$
AX \varphi \vee AX \psi \Rightarrow AX(\varphi \vee \psi)
$$

The other direction does not hold because we would need to generalize from a property that might hold for different branches to a property that has to hold for all branches.
Consider an execution with two concurrent events $e_2$ and $e_3$ where $\varphi$ holds only for $e_2$ and $\psi$ holds only for $e_3$.

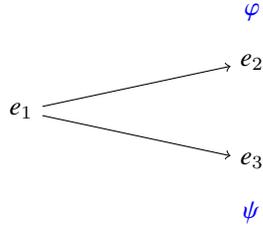
\begin{tikzpicture}
 \node (e1) {$e_1$};
 \node[above right = 0.2cm and 2.5cm of e1] (e2) {$e_2$};
 \node[below right = 0.2cm and 2.5cm of e1] (e3) {$e_3$};
 \draw[->] (e1) -- (e2);
 \draw[->] (e1) -- (e3);

 \node[above = .2cm of e2,blue] {$\varphi$};
 \node[below = .2cm of e3,blue] {$\psi$};
\end{tikzpicture}

This execution satisfies $(A, e_1)\models AX(\varphi \vee \psi)$.
Per definition, $(A,e_1)\not\models AX \varphi$ because $e_3$ does not satisfy $\varphi$ and $(A,e_1)\not\models AX \psi$ because $e_2$ does not satisfy $\psi$.
As such, we can deduce $(A,e_1)\not\models AX\varphi \vee AX \psi$ based on the semantics of $\vee$.

$$
EX (\varphi \wedge \psi) \Rightarrow EX \varphi \wedge EX \psi
$$

To see that the other direction does not hold, we can use the same argument as given above for the distributivity of $\vee$ for the $AX$ operator.
The excution above also satisfies $(A,e_1)\models EX \varphi \wedge EX \psi$ because $e_2$ satisfies $\varphi$ and $e_3$ satisfies $\psi$.
But $(A,e_1)\not\models EX (\varphi \wedge \psi)$ because we have no next event as direct successor of $e_1$ that satisfies both $\varphi$ and $\psi$.

$$
(\varphi ~U~ \psi) \vee (\varphi ~U~ \rho) \Rightarrow \varphi ~U~ (\psi \vee \rho)
$$

The other direction does not hold because we would try to deduce a stronger property about all branches based on a property that can be distributed over events on different branches.
To see this, we consider the following example:

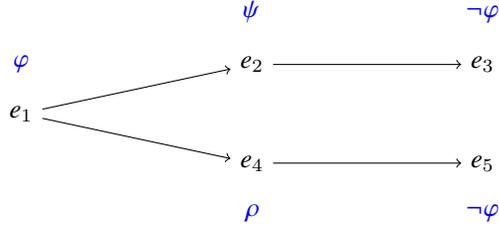
\begin{tikzpicture}
 \node (e1) {$e_1$};
 \node[above right = 0.2cm and 2.5cm of e1] (e2) {$e_2$};
 \node[right=2.5cm of e2] (e3) {$e_3$};
 \node[below right = 0.2cm and 2.5cm of e1] (e4) {$e_4$};
 \node[right=2.5cm of e4] (e5) {$e_5$};
 \draw[->] (e1) -- (e2);
 \draw[->] (e2) -- (e3);
 \draw[->] (e1) -- (e4);
 \draw[->] (e4) -- (e5);

 \node[above = .2cm of e1,blue] {$\varphi$};
 \node[above = .2cm of e2,blue] {$\psi$};
 \node[above = .2cm of e3,blue] {$\neg\varphi$};
 \node[below = .2cm of e4,blue] {$\rho$};
 \node[below = .2cm of e5,blue] {$\neg\varphi$};
\end{tikzpicture}

Event $e_1$ satisfies $\varphi$.
Branching of after $e_1$, we have two concurrent event strands $e_2$ to $e_3$ and $e_4$ to $e_5$.
Event $e_2$ satisfies $\psi$, so the execution only consisting of $e_1$ to $e_3$ would satisfy $\varphi ~U~ \psi$.
Event $e_4$ satisfies $\rho$, so the execution only consisting of $e_1$ to $e_5$ would satisfy $\varphi ~U~ \rho$.
Hence, $(A, e_1)\models \varphi ~U~ (\psi \vee \rho)$.
Even though the individual executions sketched above satisfy the properties, it holds that $(A,e_1)\not\models\varphi ~U~ \psi$ as well as $(A,e_1)\not\models \varphi ~U~ \rho$.
The reason is that we have to consider the other branch as well.
We have $(A,e_1)\not\models \varphi ~U~ \psi$ because $e_5$ does not satisfy $\varphi$ and thus we require an event in the chain of events between $e_1$ and $e_5$ that satisfies $\psi$.
But there is no such event.
The same reasoning can be applied for $(A,e_1)\not\models \varphi ~U~ \rho$.

\paragraph{Negation.}
The usual equalities to reason about negation hold for EPTL.
\begin{eqnarray*}
 \neg(EX \varphi) &\equiv& AX(\neg \varphi) \\
 \neg(AX \varphi) &\equiv& EX(\neg \varphi) \\
 \neg(F \varphi) &\equiv& G(\neg \varphi)   \\
 \neg(G \varphi) &\equiv& F (\neg \varphi)
\end{eqnarray*}

It is not very surprising that the negation of an existential step $EX$ is an universal step $AX$; similarly for $F$ and $G$ as $F$ has existential qualities whereas $G$ has universal qualities.

There is just one rule which surprisingly does not hold for EPTL: $AX \varphi \not \Rightarrow EX \varphi$.
The reason for this unexpected behavior is that we want to consider not only infinite but also finite executions.
In fact, we can show that for a last event of an abstract execution it holds that:

\begin{eqnarray*}
 \lastevent(e, A) &\Rightarrow& (A,e)\models AX\varphi                                             \\
 \lastevent(e, A) &\Rightarrow& (A,e)\models \neg EX \varphi                                       \\
 \neg \lastevent(e, A) &\Rightarrow& ((A,e)\models AX \varphi \Rightarrow (A,e)\models EX \varphi)
\end{eqnarray*}

In an abstract execution, there can be multiple last events; these are essentially all events that have no successor event.
These events are special in that $AX \varphi$ holds for every $\varphi$, especially $AX ~\mathit{false}$.
On the other hand, $\neg EX \varphi$ holds for every $\varphi$, especially $\neg EX \mathit{true}$.
These properties have already been observed by \citet{havelund_testing_2001} and \citet{de_giacomo_reasoning_2014}.

\paragraph{Idempotence.}
Next, we have proven that idempotence holds for all operators introduced in EPTL.

\begin{eqnarray*}
 F (F \varphi) &\equiv& F \varphi                         \\
 G (G \varphi) &\equiv& G \varphi                         \\
 \varphi ~U~ (\varphi ~U~ \psi) &\equiv& \varphi ~U~ \psi
\end{eqnarray*}

The rules allow to remove unnecessary operators when reasoning about the validity of a formula.

\paragraph{Induction.}
The last set of rules deals with reasoning about the validity of formulas in general.
One typical approach is to use induction on the events of the abstract execution.
Indeed, the induction formulas for the $F$ and $G$ operators hold:

\begin{eqnarray*}
 F \varphi &\equiv& \varphi \vee EX(F \varphi)   \\
 G \varphi &\equiv& \varphi \wedge AX(G \varphi)
\end{eqnarray*}

Unfortunately, it remains unclear whether an induction formula for the $U$ operator exists.
This makes reasoning about formulas including this operation harder because the reasoning has to be done solely based on the abstract execution.

The induction formula for $U$ in LTL is $\varphi ~U~ \psi \equiv \psi \vee (\varphi \wedge X(\varphi ~U~ \psi))$.
In EPTL we have a different set of operators.
We checked both using an $AX$ and an $EX$ operator instead of the $X$ of LTL;
both formulas can be shown to be invalid.
Especially the direction from left to right is interesting regarding the application to partial orders.

We consider an abstract execution with the following events:

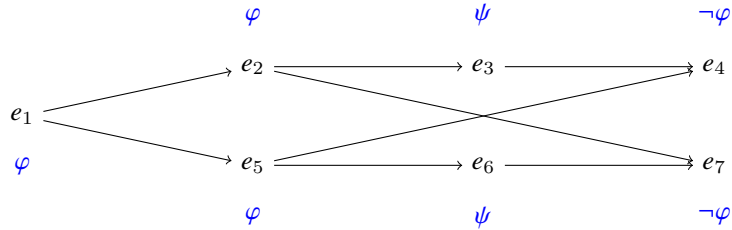
\begin{tikzpicture}
  \node (e1) {$e_1$};
  \node[above right = 0.2cm and 2.5cm of e1] (e2) {$e_2$};
  \node[right = 2.5cm of e2] (e3) {$e_3$};
  \node[right = 2.5cm of e3] (e4) {$e_4$};
  \node[below right = 0.2cm and 2.5cm of e1] (e5) {$e_5$};
  \node[right = 2.5cm of e5] (e6) {$e_6$};
  \node[right = 2.5cm of e6] (e7) {$e_7$};
  \draw[->] (e1) -- (e2);
  \draw[->] (e2) -- (e3);
  \draw[->] (e3) -- (e4);
  \draw[->] (e1) -- (e5);
  \draw[->] (e5) -- (e6);
  \draw[->] (e6) -- (e7);
  \draw[->] (e5) -- (e4);
  \draw[->] (e2) -- (e7);

 \node[below = .2cm of e1,blue] {$\varphi$};
 \node[above = .2cm of e2,blue] {$\varphi$};
 \node[above = .2cm of e3,blue] {$\psi$};
 \node[above = .2cm of e4,blue] {$\neg\varphi$};
 \node[below = .2cm of e5,blue] {$\varphi$};
 \node[below = .2cm of e6,blue] {$\psi$};
 \node[below = .2cm of e7,blue] {$\neg\varphi$};
\end{tikzpicture}

We have two concurrent executions $e_2$ to $e_4$ and $e_5$ to $e_7$ where $\varphi ~U~ \psi$ holds independently for both $e_2$ and $e_5$.
The event $e_1$ is visible to both concurrent executions.
Additionally we have an intermediate synchronization such that $e_2$ is visible to the other process before execution of $e_7$ and $e_5$ is visible before execution of $e_4$.
From this construction we can convince ourself that $(A, e_1)\models \varphi ~U~ \psi$.

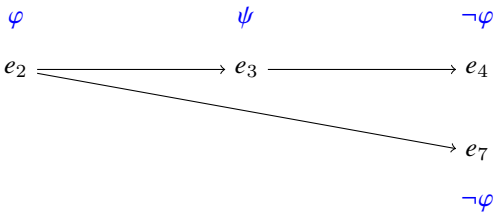
\begin{tikzpicture}
  \node (e1) {};
  \node[above right = 0.2cm and 2.5cm of e1] (e2) {$e_2$};
  \node[right = 2.5cm of e2] (e3) {$e_3$};
  \node[right = 2.5cm of e3] (e4) {$e_4$};
  \node[below right = 0.2cm and 2.5cm of e1] (e5) {\phantom{$e_5$}};
  \node[right = 2.5cm of e5] (e6) {\phantom{$e_6$}};
  \node[right = 2.5cm of e6] (e7) {$e_7$};
  \draw[->] (e2) -- (e3);
  \draw[->] (e3) -- (e4);
  \draw[->] (e2) -- (e7);

  \node[above = .2cm of e2,blue] {$\varphi$};
  \node[above = .2cm of e3,blue] {$\psi$};
  \node[above = .2cm of e4,blue] {$\neg\varphi$};
  \node[below = .2cm of e7,blue] {$\neg\varphi$};
\end{tikzpicture}

Looking just at the subgraph starting with $e_2$, the picture changes.
The execution $e_2$ through $e_4$ satisfies $\varphi ~U~ \psi$ by construction.
Event $e_7$ also happens after $e_2$ and does not satisfy $\varphi$.
But $e_2$ does also not satisfy $\psi$ which means that $(A, e_2)\not\models \varphi ~U~ \psi$.
The same reasoning can be applied to the subgraph starting with $e_5$.
From this we can deduce that $(A, e_1)\not\models EX(\varphi ~U~ \psi)$ and $(A,e_1)\not\models AX(\varphi ~U~ \psi)$.
The induction formula for $U$ translated from LTL is therefore not a rule for EPTL.

\section{Verification of Implementations}

In previous work \citep{weber_access_2016}, we have shown how to implement access control in weakly consistent systems.
But the specification of the correctness criterium is informal and the model of the implementation cannot be checked to satisfy this criterium.
Using EPTL, we can specify and formally verify the correctness of the implementation model of \citep{weber_access_2016}.
We have modeled EPTL in the theorem prover Isabelle/HOL.
All laws of EPTL are formalized and verified in the interactive theorem prover and are used by the tool to simplify formulas.
Even though we did not yet find an efficient automatic checking procedure for EPTL, the proofs can be done in semi-automatic fashion in HOL.
Together with the relatively strong automation of Isabelle/HOL this should make for a comfortable environment in which to show that the presented model is suitable to implement access control.

\section{Related Work}

Partial order semantics has been used before as an intuitive representation of the execution of concurrent systems.
But it is assumed that the semantics does not need to distinguish among total-order executions that are equivalent up to reordering of some class of events.
The notion of independent events is not easily defined in a weakly consistent setting.

\citet{alur_deciding_2005} presented a global partial order logic called ISTL.
Same as we, they do not restrict the view on the system to the state sequence observed by a local process.
The logic is based on a partially ordered set of local states which can also be seen as a branching structure.
This branching structure represents all possible sequences of global states that may be derived from the partial order.
This state based approach makes it unsuitable for reasoning about weakly consistent systems.
As described in Section \ref{sec:LTLnotsuitable}, encoding the events and the conflict resolution strategy into a state requires knowledge about the implementation of the conflict resolution strategy.
Since the concrete implementation has to be abstracted from in the specification of the behavior of a weakly consistent system, ISTL is not suitable as a specification language for weakly consistent systems.

The other line of research about partial order semantics uses Mazurkiewicz traces \citep{mazurkiewicz_concurrent_1977}.
The base for these traces is a finite set of actions, which can be seen as state transformations of resources of the system under investigation.
Two actions are independent if they act on disjoint set of resources.
Only independent actions are allowed to be performed concurrently.
This restriction is the reason why Mazurkiewicz traces cannot be used to reason about weakly consistent systems in the given form.
In these considered systems, the resources are shared objects where each process has an own copy of the object called a replica.
Actions or operations on these objects are performed on this local copy without synchronization, the resulting conflicts are resolved when synchronizing the state changes between different replicas.
When looking at these operation from a global view, they all change the same shared object.
In this sense, the operations are not independent, even though they are possibly performed concurrently.
It is not obvious how to apply Mazurkiewicz traces to weakly consistent systems.

A common way to specify properties about weakly consistent systems is to directly specify them based on the abstract execution.
\citet{gotsman_composite_2015} as well as \citet{zeller_towards_2016} both use invariants about the execution which are extended by properties about the visibility between events.
This makes the specification less readable because the temporal aspects of the specification are mixed with properties about the actual events.
By separating the temporal aspects, EPTL specifications are closer to the intuitive natural language formulation of the required properties.

\section{Conclusion and Future Work}

We presented the new temporal logic EPTL that is tailored to specify properties of weakly consistent systems.
The specifications are based on the global partial order between events in a replicated system.
The complete logic is modeled in Isabelle/HOL and all laws are verified using the theorem prover.
All theory files are available under \url{https://softech-git.informatik.uni-kl.de/mweber/EPTL/tree/master}.

With only the given future fragment of EPTL, it is not possible to express the complete semantics of CRDTs\citep{crdt,burckhardt2014} like the multi-value register.
Adding a past fragment with a since-operator will enable to use EPTL as such a specification language.

\bibliographystyle{splncsnat}
\bibliography{references}

\end{document}